\documentclass[aps,twocolumn,superscriptaddress,prl]{revtex4-1}

\usepackage{graphicx}

\begin{document}
\title{Thermoelectric determination of electronic entropy change in Ni-doped FeRh}

\author{Nicol\'as P\'erez}
\affiliation{IFW Dresden, Institute for Metallic Materials, Helmholtzstrasse 
20, 01069 Dresden, Germany}

\author{Alisa Chirkova}
\affiliation{IFW Dresden, Institute for Metallic Materials, Helmholtzstrasse 
20, 01069 Dresden, Germany}
\affiliation{TU Dresden, Institute of Materials Science, 01062, Dresden, 
Germany}

\author{Konstantin P. Skokov}
\affiliation{Materials Science, TU Darmstadt, 64287, Darmstadt, Germany}
\affiliation{National University of Science and Technology, 119049, 
Moscow, Russia}

\author{Thomas George Woodcock}
\affiliation{IFW Dresden, Institute for Metallic Materials, Helmholtzstrasse 
20, 01069 Dresden, Germany}

\author{Oliver Gutfleisch}
\affiliation{Materials Science, TU Darmstadt, 64287, Darmstadt, Germany}

\author{Nikolai V. Baranov}
\affiliation{Institute of Natural Sciences and Mathematics, Ural Federal 
University, Yekaterinburg, 620083, Russia}
\affiliation{M. N. Miheev Institute of Metal Physics, Ural Branch of Russian 
Academy of Sciences, Yekaterinburg, 620990, Russia}

\author{Kornelius Nielsch}
\affiliation{IFW Dresden, Institute for Metallic Materials, Helmholtzstrasse 
20, 01069 Dresden, Germany}
\affiliation{TU Dresden, Institute of Materials Science, 01062, Dresden, 
Germany}

\author{Gabi Schierning}
\affiliation{IFW Dresden, Institute for Metallic Materials, Helmholtzstrasse 
20, 01069 Dresden, Germany}

\begin{abstract}
The net entropy change corresponding to the charge carriers in a Ni-doped FeRh bulk polycrystal was experimentally evaluated in a single sample using low temperature heat capacity experiments with applied magnetic field, and using Seebeck effect and Hall coefficient measurements at high temperatures across the first order transition. From the heat capacity data a value for the electronic entropy change \(\Delta S_{el}\approx8.9\) J\ kg\(^{-1}\)K\(^{-1}\) was extracted, whereas a value of up to 4  J\ kg\(^{-1}\)K\(^{-1}\) was obtained form the Seebeck coefficient. Additionally, the analysis of the Seebeck coefficient allows tracing the evolution of the electronic entropy change with applied magnetic field. An increase of the electronic entropy with increasing applied magnetic field is evidenced, as high as 10 percent at 6 T.
\end{abstract}

\maketitle
The determination of entropy changes in solids at phase transitions of different kinds is of major relevance for understanding fundamental phenomena in  materials, and a source of information for their design and optimization for applied purposes. Typical examples can be found in the phase transitions of magnetocaloric materials \cite{Casanova2005}, shape memory alloys \cite{Manosa2010}, or piezo/ferroelectrics \cite{Zhang2016}. The total entropy change is the result of several contributions coming from the ion lattice, magnetization changes, electrical polarization changes, or conduction electrons. Evaluating the different contributions is necessary to understand the transition processes. Particularly, the entropy associated with the conduction electrons is crucial in the understanding of correlated electron systems \cite{Canfield2016}, including magnetic \cite{Hagymasi2016}, semiconducting \cite{Lu2017}, and superconducting \cite{Hirsch2017} materials. In that context, the experimental determination of thermodynamical quantities in low dimensional systems is a matter of current interest \cite{Pudalov2015}.

Here we investigate the first-order metamagnetic transition in Ni-doped \(\alpha\)-FeRh. The transition from a low temperature antiferromagnetic (AF) phase to a high temperature ferromagnetic (FM) phase \cite{Tu1969, Baranov1995, Kreiner1998}, shows a large entropy change reaching \(\Delta S\approx 16\) J kg\(^{-1}\)K\(^{-1}\;\) \cite{Chirkova2016}, and a change in lattice parameter \cite{DeBergevin1961, Ibarra1994}. Different works have attributed the origin of the transition to processes involving conduction electrons \cite{Tu1969, Baranov1995}, magnetic instability associated with magnon modes \cite{Gu2005}, or lattice instability \cite{Moruzzi1992}. The different entropy contributions to the phase transition have been evaluated either theoretically \cite{Wolloch2016}, or experimentally using a set of proxi thin film samples for the different magnetically ordered states \cite{Cooke2012}. Recent theoretical studies addressed the importance of changes in electronic structure at the phase transition \cite{Mankovsky2017}, and an additional barrierless martensitic transition has been predicted to occur below about 90 K \cite{Zarkevich2018}.

In this work we have studied the electronic part of that singular phase transition in a Ni-doped FeRh polycrystal using low temperature heat capacity. Additionally, using Seebeck and Hall coefficient measurements, the evolution of the entropy change with applied field was evaluated. The differences between calorimetric and electronic transport approaches are discussed. The Hall coefficient measurements indicate a complex magnetic behaviour in the AF phase before the transition, and seem to agree with the recently theoretically predicted existence of an additional transition at approximately 90 K. 

\subsection{Evaluation of the electronic entropy}
In the usual experimental approach, heat capacity measurements at low temperatures  are performed and interpreted making use of the approximated series expansion \(C_p=\gamma\;T+\sum_{i=3,5,7\dots}\beta_i\;T^i+C_m.\label{heatcap}\), where \(\gamma\) is the electronic specific heat coefficient, \(\beta_i\) are the lattice specific heat coefficients, and \(C_m\) is the magnetic contribution to the specific heat. However, in the presence of a phase transition at higher temperatures, typically several samples of different compositions, and different magnetic ordering at low temperature, need to be studied in order to experimentally determine difference in \(\gamma\) between the differently magnetically ordered states \cite{Kreiner1998, Cooke2012, Tu1969}. In our approach, low temperature \(C_p\) was measured in the same sample for the different magnetically ordered states forcing the metamagnetic transition with an applied magnetic field of sufficient intensity. The Seebeck coefficient \(\alpha\) inherently contains information about the entropy of the conduction electrons, as derived by Ioffe \cite{Ioffe1957} and recently by Goupil \cite{Goupil2011}, both from thermodynamical arguments. A simplified derivation can be performed for isotropic materials using the definition of the Peltier coefficient, \(\Pi=Q/(e\,N_t)\), as the heat \(Q\) per transported carrier \(N_t\), and Onsager's reciprocal relation, \(\Pi=\alpha\cdot T\). Both relations lead in a simple way to the formal expression that relates \(\alpha\) and  the entropy associated with the effectively transported charge carriers, \(\alpha=Q/(T\,e\,N_t) = S_t/(e\,N_t)\). The value of the elemental charge \(e\) is conventionally taken with sign depending on the type of carrier. It is worth noting that this thermodynamical expression of \(\alpha\) is of general applicability, and is also valid for materials with noticeable electronic correlations or in which conduction does not take place near the Fermi energy. Nevertheless microscopic models can be built, such as Mott's formula that account for the measured \(\alpha\) \cite{Mott1936}. This approach delivered good agreement with experimental results  relating \(\alpha\) to the electronic transport and magneto transport properties of some materials (see \cite{Frauen2015} and references therein).

\subsection{Experimental results}
The experiments were performed on a bulk polycrystalline sample of composition (Fe\(_{0.96}\)Ni\(_{0.02}\))Rh\(_{1.02}\) cut in rod shape with rectangular cross section. A thin slice cut off the rod was used for \(C_p\) experiments. X-ray diffraction evidenced the well-ordered CsCl-type phase with the presence of the paramagnetic fcc phase. By means of scanning electron microscopy the amount of the fcc phase was estimated to be approximately 6 vol.\%. Detailed description of the synthesis procedure can be found elsewhere \cite{Baranov1995}.  Hall coefficient, magnetization, \(\alpha\) and \(C_p\) measurements under applied magnetic field were performed with Quantum Design Physical Property Measurement Systems (PPMS).

\begin{figure}[htbp]
\begin{center}
\includegraphics[width=8cm]{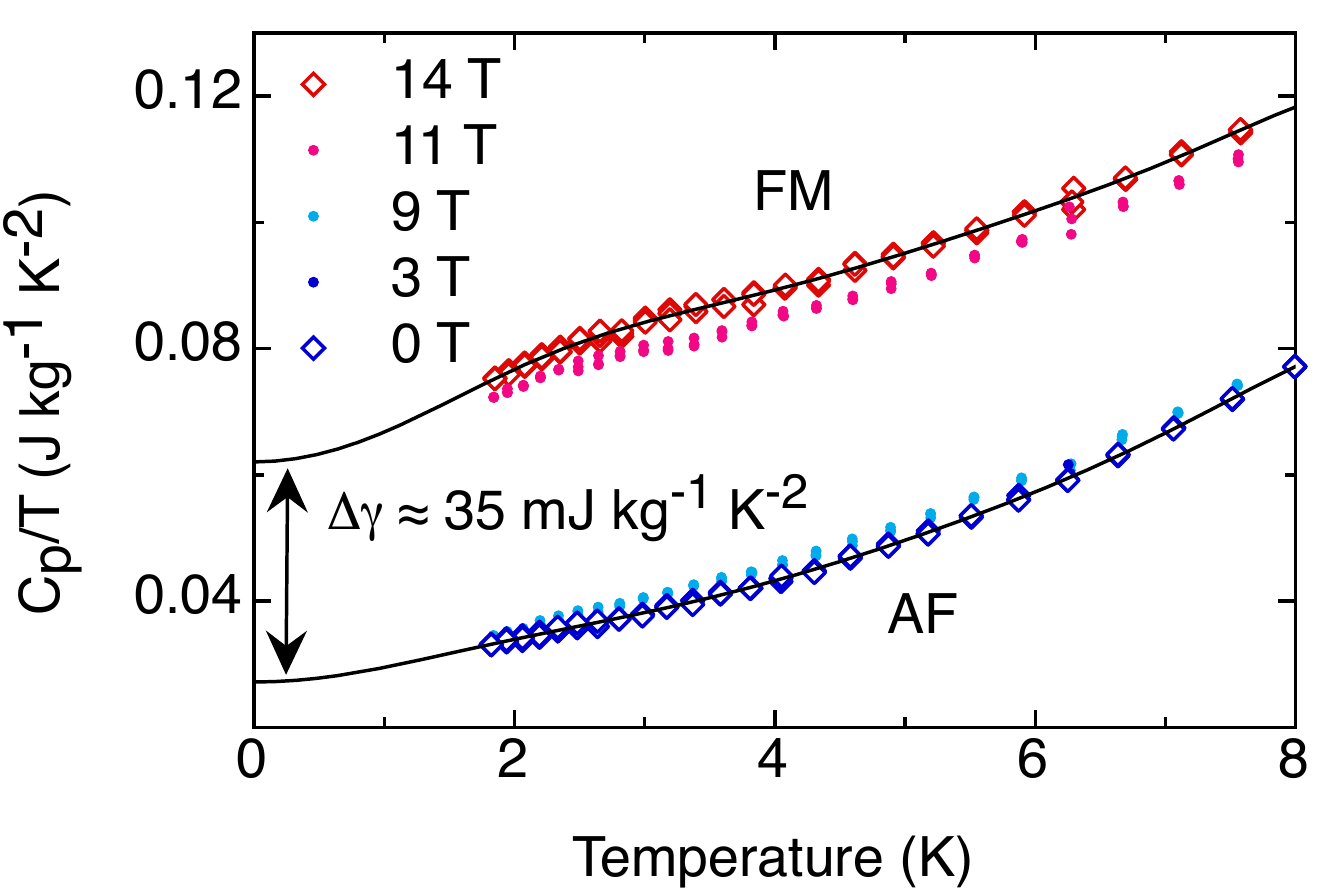}
\caption{Low temperature heat capacity data for Ni-doped FeRh with different applied magnetic fields. At a field of 11 T and higher the system is in the FM state. Solid lines are fitted curves.}
\label{figCp}
\end{center}
\end{figure}

\begin{figure}[t]
	\centering\mdseries
		\includegraphics[width=8 cm]{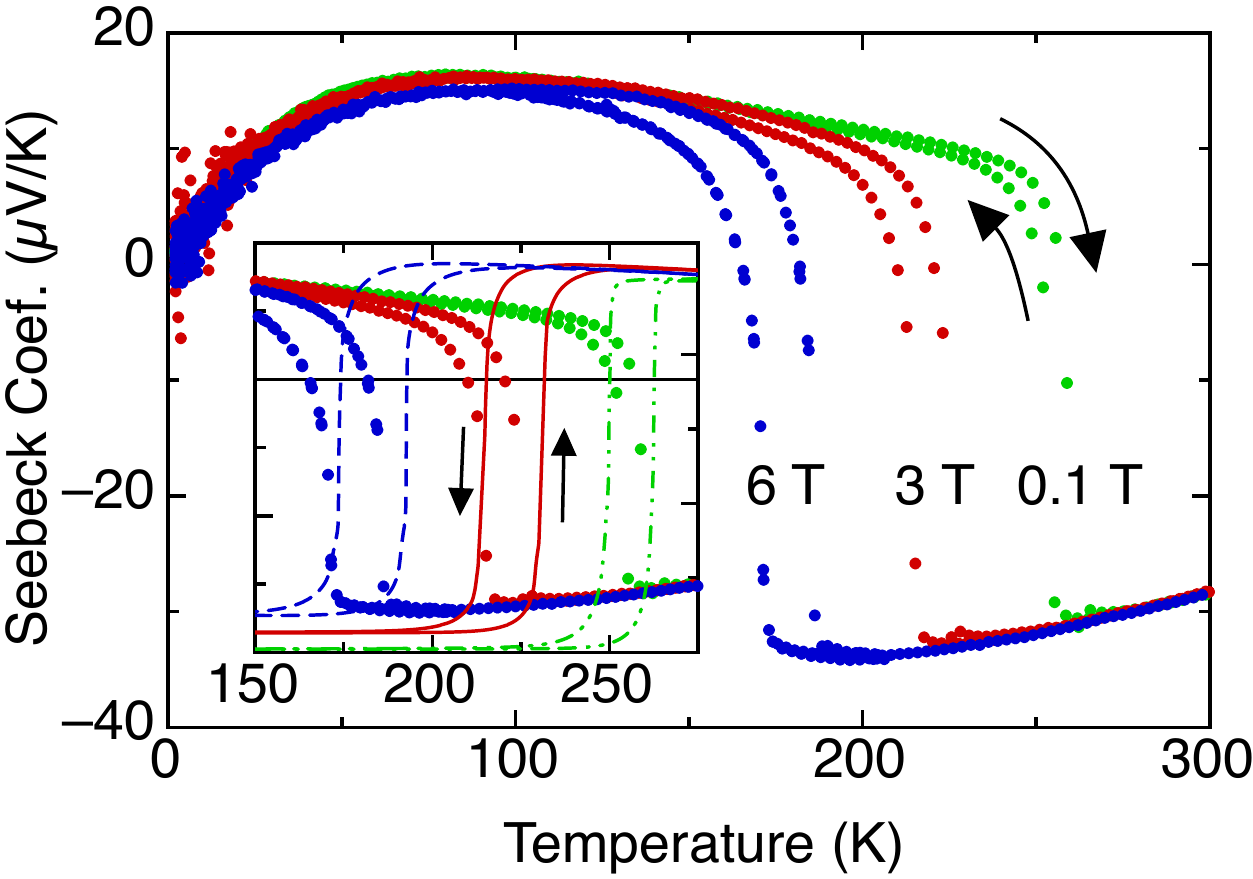}
	\caption{Measured values of \(\alpha\) acquired at different magnetic fields. Inset: Comparison of \(\alpha\) with temperature dependant magnetization curves (lines)}
	\label{figSeebeck}
\end{figure}

Figure \ref{figCp} shows the low temperature data for \(C_p\), where the the transition to the FM state at magnetic fields of 11 T and higher can be identifed. An anomaly can be seen at the lowest temperatures, which in other magnetic systems has been attributed to a gap \(\Delta\) in the magnon spectrum \cite{Majzlan2003}. For that reason, the magnetic contribution to \(C_p\) had the form \(C_m = B_{FM}T^{3/2}e^{(-\Delta/T)}\) for the ferromagnetic phase and \(C_m = B_{AF}T^{3}e^{(-\Delta/T)}\) for the antiferromagnetic phase \cite{Majzlan2003}. The fitted values for \(\gamma\) were 62\(\pm\)2 mJ\,kg\(^{-1}\)\,K\(^{-2}\), and 27\(\pm\)1 mJ\,kg\(^{-1}\)\,K\(^{-2}\) for the FM and AF phases respectively. Extrapolating the difference in \(\gamma\) to transition temperature, \(T_t\), which at zero field is 256 K, results in an electronic entropy change \(\Delta S_{el} = \Delta\gamma\;T_t = 8.9\pm0.2\) J\ kg\(^{-1}\)K\(^{-1}\). This result is in good agreement with the value reported previously by Cooke \emph{et. al.} using different thin film proxies for the FM and AF phases \cite{Cooke2012}.

In Figure \ref{figSeebeck} the measured values of \(\alpha\) are plotted, showing a jump and a change of sign at the expected transition temperatures. The shift in temperature of the jump is accompanied by an increased hysteresis in the AF region. In that region the 6 T curve shows some slight deviation form the 0.1 ands 3 T curves, which superimpose almost perfectly. In the FM phase, the heating and cooling curves superimpose perfectly within the experimental error for the three fields examined. Our data are in good agreement with the results obtained by Kobayashi \emph{et al.} in Ni-doped FeRh \cite{Kobayashi2001}. In the FM phase the applied magnetic field does not change the values of \(\alpha\). The displacement of the jump in \(\alpha\) agrees with the one in magnetization (Fig. \ref{figSeebeck} inset). At higher fields and at temperatures lower than \(T_t\), a residual non zero magnetization is observed, indicating the presence of some ferromagnetically ordered part of the material.

\begin{figure}[t]
	\centering
		\includegraphics[width=8 cm]{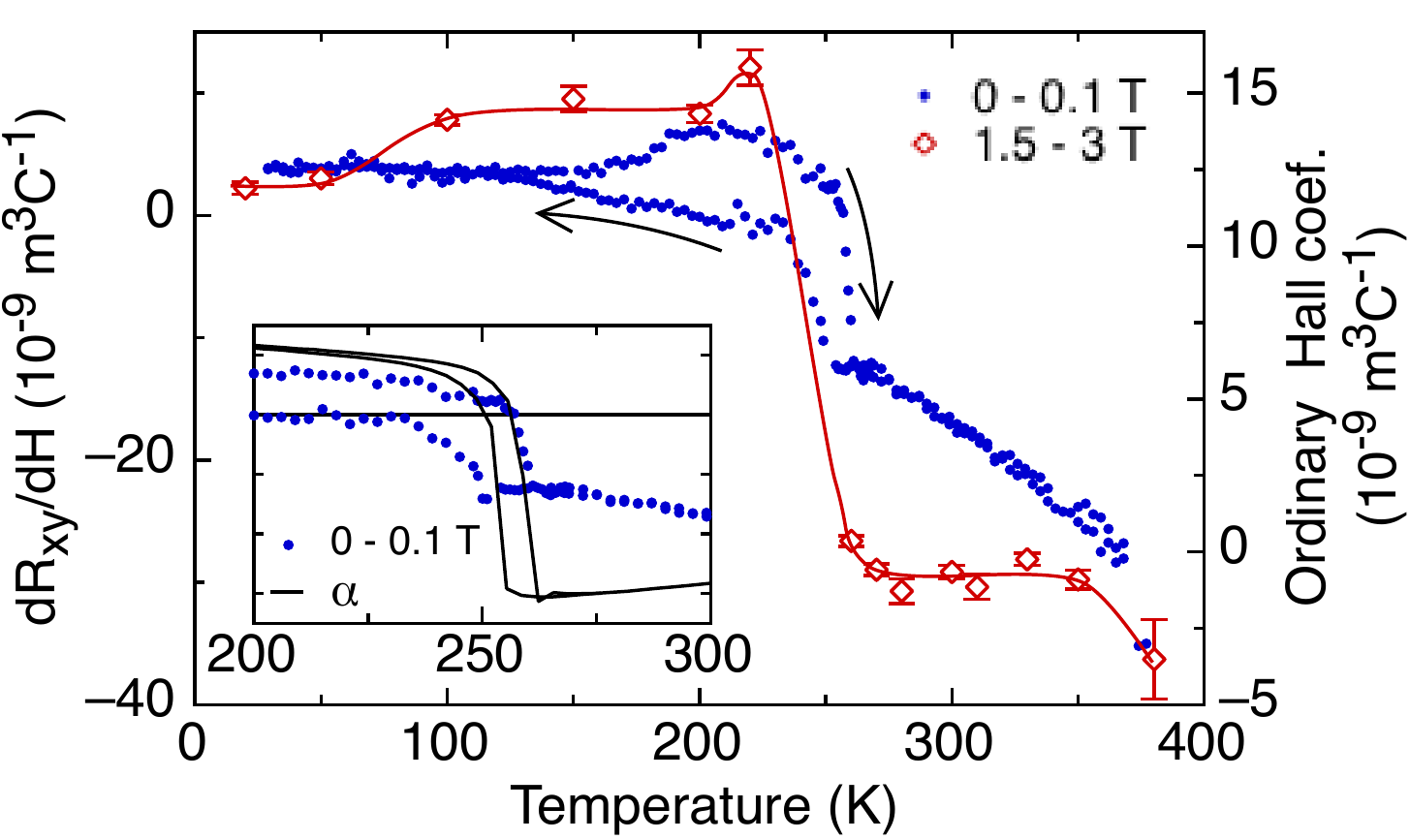}
	\caption{\(dR_{xy}/dH\) measured at low fields and at magnetic saturation (ordinary Hall coefficient). Line is a guide to the eye. Inset: Detail of the low field \(dR_{xy}/dH\) and \(\alpha\) at the transition region. Horizontal line is the zero level for both data sets.}
	\label{figHall}
\end{figure}

Transverse resistance, \(R_{xy}\), was measured as a function of applied magnetic field in one experiment up to 0.1 T and in a second experiment with applied field of up to 3 T. The temperature was changed with the sample in a demagnetized condition. The low field slope \(dR_{xy}/dH\) contains additional information about the initial susceptibility of the sample. Form the high field slope, at fields higher than about 1.5 T when the sample is magnetically saturated, the ordinary Hall coefficient, \(r_h\), can be directly obtained \cite{Nagaosa2010}. In Fig. \ref{figHall} a significant contribution arising from the anomalous Hall coefficient can be clearly identified in  \(dR_{xy}/dH\) for the high temperature FM phase. An approximately linear decrease of  \(dR_{xy}/dH\) is observed as the temperature decreases towards the transition. A plateau is observed at temperatures just above \(T_t\) both, on cooling and on heating. Additionally, a complex hysteresis is seen in \(dR_{xy}/dH\) in the AF phase down to approx.\ 120\ K, which contrast with the hysteresis observed in \(\alpha\), which shows no additional features in the AFM region (\ref{figHall} inset). A nearly constant \(r_h\) is obtained before and after the transition, with a jump and sign change in the transition region. Moreover, no hysteresis was evidenced in \(r_h\) in the temperature range 120-260\ K. The low value of \(r_h\) in the FM phase coincides with the observations of deVries \emph{et al.} in FeRh thin films \cite{DeVries2013}, and indicates the contribution of valence and conduction bands to the electronic transport, resulting in a situation close to compensation. Additionally, \(r_h\) changes abruptly between 100 and 50 K, which could be related to a recently predicted martensitic transformation occurring at about 90 K \cite{Zarkevich2018}. 

Mangetization and transverse resistance point out that local FM spin arrangements occur in the AF phase. The \(dR_{xy}/dH\) indicate a complex magnetic interactions in the AF phase starting from 120\ K until \(T_t\). In this respect, it has been recently shown that the presence of the fcc phase influences the transition temperature and hysteresis \cite{Chirkova2017}. The actual way in which these possible local spin arrangements may be realized, the ferromagnetic background seen in the AF phase, and the influence of the residual paramagnetic fcc phase present in the sample are subjects outside the scope of this article. 
\begin{figure}[tb]
\begin{center}
\includegraphics[width=8 cm]{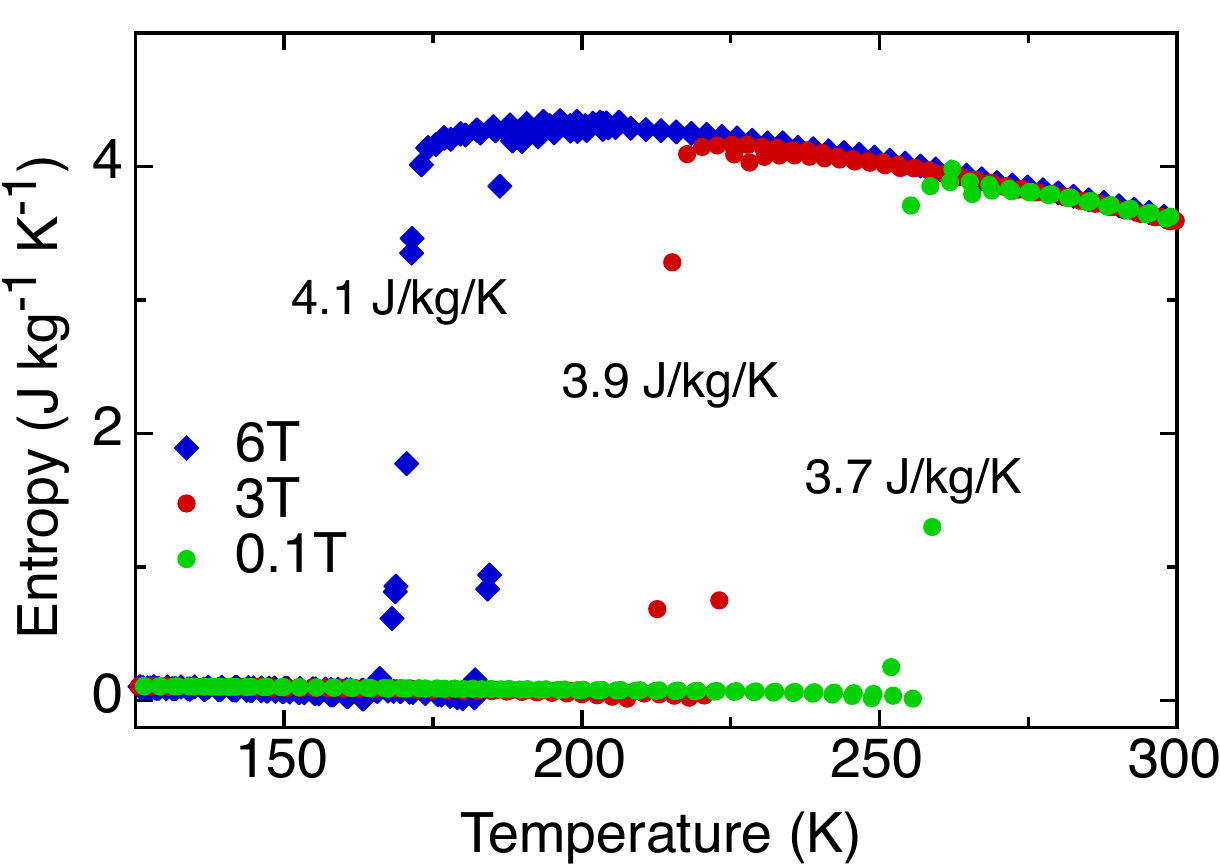}
\caption{Evolution of \(S_t\) with temperature for different magnetic fields.}
\label{figEntropy}
\end{center}
\end{figure}

\subsection{Discussion}
According to the relation \(\alpha= S_t/(e\,N_t)\), the effective transported charge needs to be known in order to extract the entropy information from the Seebeck coefficient. An effective carrier concentration was inferred from \(r_h\), and a measure of entropy associated with the transported charge was calculated as \(S_{t}=\alpha/r_h\) (Fig.\ref{figEntropy}). The values obtained in this way have to be regarded as an approximation within Drude transport model. In the case of multi band conduction, \(r_h\) and \(\alpha\) are typically regarded as conductivity weighted averages over the contributing bands, different for each coefficient \cite{Putley1975}. However, according to the data in Fig. \ref{figHall}, it is safe to use \(r_h\) to obtain a constant effective carrier concentration before and after the transition region. The entropy values obtained in this way will deviate from the actual ones by a constant factor, which we cannot determined with the available experimental data. However, information on the evolution of the electronic entropy and of the entropy change can be obtained. The estimated entropy change increases with the applied magnetic field ranging form \(\Delta S_t\approx3.7\) J kg\(^{-1}\)K\(^{-1}\) at 0.1 T to \(\Delta S_t\approx4.1\) J kg\(^{-1}\)K\(^{-1}\) at 6 T (Fig.\ref{figEntropy}). For the reasons mentioned previously, these results are approx. a factor 2 smaller than the one obtained from \(C_p\). However, it is clear that the electronic entropy change does not remain constant with applied magnetic field. This evidences an intrinsic shortcoming of the approach to characterize the entropy associated to the transported charge using \(C_p\) measurements at different magnetic fields or using proxis of different magnetically ordered states. The relative increase in the electronic entropy change with the applied magnetic field would then be of about 10\% between zero and 6 T. 

Note that in the FM state the applied magnetic field does not change the the value of \(\alpha\) at a given temperature outside the transition region. Instead, the magnetic field keeps the material in the FM state until lower temperatures allowing \(\alpha\) to monotonically become more negative. Consequently, an increase in entropy is evident in the FM phase when cooling down to \(T_t\). And the obtained \(\Delta S_t\) increases with the applied magnetic field because \(\alpha\) increases in the FM state at the corresponding \(T_t\). In the AF phase, the variation of the entropy with temperature is rather small. However, some tendency to decrease the entropy with increasing temperature up to the transition may be noticed in the evolution of \(\alpha\) and also in the increase of \(r_h\) (Fig. \ref{figHall}). This indicates that the difference in the electronic entropy between the AF and FM phases increases with approaching \(T_t\) from both sides. Such a behaviour might be attributed to the enhancement of the spin fluctuation contribution in the d electron subsystem in the vicinity of the magnetic instability. In fact, the AF-FM transition in FeRh is accompanied by the changes in the Rh moment from zero up to approximately 1 \(\mu_B\), while the magnetic moment on Fe atoms does not remarkably vary (see \cite{Lewis2016} and references therein). Bearing that in mind, a spin-fluctuation contribution may be mainly associated with the Rh sublattice \cite{Ju2004}. The role of spin fluctuations together with the difference in the density of electronic states at the AF and FM order in the formation of properties of FeRh was recently demonstrated with first-principles calculations in the frame of the density functional theory \cite{Mankovsky2017}.

\subsection*{Conclusions}
The electronic entropy of a Ni-doped FeRh polycrystal was determined using low temperature \(C_p\) measurements with applied magnetic fields. Making use of the thermodynamical definition of the Seebeck coefficient, in combination with Hall coefficient measurements allowed us to give a direct measure of the entropy change associated with the conduction electrons across the first order metamagnetic phase transition. The values obtained with the two methods differed by a factor 2. However, it was demonstrated that the applied magnetic field causes the electronic entropy change to increase up to 10\% between 0.1 T and 6 T.

\subsection*{Acknowledgements}
N.V.Baranov acknowledges the support of the FASO of Russia through project No. 01201463328. The authors thank Dr. Sebastian F\"ahler for insightful discussions.

\end{document}